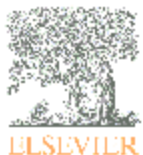

Journal logo

# Excitation Functions of Product Nuclei from 40-2600 MeV Proton-Irradiated $^{206,207,208,nat}$Pb and $^{209}$Bi

Yury E. Titarenko,[a,*] Viacheslav F. Batyaev,[a] Ruslan D. Mulambetov,[a] Valery M. Zhivun,[a] Vladilen S. Barashenkov,[b] Stepan G. Mashnik,[c] Yury N. Shubin,[d] Anatoly V. Ignatyuk[d]

[a] *Institute for Theoretical and Experimental Physcis, 117218 Moscow, Russia*

[b] *Joint Institute for Nuclear Research, 141980 Dubna, Russia*

[c] *Los Alamos National Laboratory, Los Alamos, NM 87545, USA*

[d] *Institute for Physics and Power Engineering, 249020 Obninsk, Russia*

**Abstract**

The work is aimed at experimental determination of the independent and cumulative yields of radioactive residual nuclei produced in intermediate-energy proton-irradiated thin targets made of highly isotopic enriched and natural lead ($^{206,207,208,nat}$Pb) and $^{209}$Bi. 5972 radioactive product nuclide yields have been measured in 55 thin targets induced by 0.04, 0.07, 0.10, 0.15, 0.25, 0.6, 0.8, 1.2, 1.4, 1.6, and 2.6 GeV protons extracted from the ITEP U-10 proton synchrotron. The measured data have been compared with data obtained at other laboratories as well as with theoretical simulations by seven codes. We found that the predictive power of the tested codes is different but is satisfactory for most of the nuclides in the spallation region, though none of the codes agree well with the data in the whole mass region of product nuclides and all should be improved further. © 2001 Elsevier Science. All rights reserved nuclear reaction; spallation; fission; fragmentation; yields; residual nuclides; cross sections; simulation; Monte-Carlo codes; comparison.

## 1. Introduction

A number of current and planned nuclear projects, such as transmutation of nuclear wastes with Accelerator-Driven Systems (ADS), require a large amount of nuclear data. Since not all the required data can be measured, reliable models and codes are to be used in such projects. The codes should be such projects. The codes should be verified, validated, and benchmarked against measurements that are as reliable as possible.

During 2002-2004, under the ISTC Project # 2002 [1], ITEP has realized an experimental program to measure the residual nuclide production cross sections in $^{208, 207, 206}$Pb, $^{nat}$Pb, and $^{209}$Bi thin targets irradiated with protons of 0.04, 0.07, 0.10, 0.15, 0.25, 0.4, 0.6, 0.8, 1.2, 1.6, and 2.6 GeV. In the present work, we present part of our data and analyze all our measured cross sections with seven codes used in many current applications in order to validate their predictive powers.

## 2. Experiment

The thin $^{208, 207, 206, nat}$Pb and $^{209}$Bi targets of 10.5 mm diameter (127 - 358 mg/cm$^2$ thickness) together with aluminum monitors of the same diameter (127 – 254 mg/cm$^2$ thickness) were irradiated using the external beam of the ITEP U10 proton synchrotron. The targets used were of the following isotopic composition: $^{208}$Pb (0.87%$^{206}$Pb, 1.93%$^{207}$Pb, 97.2%$^{208}$Pb); $^{207}$Pb (0.03%$^{204}$Pb, 2.61%$^{206}$Pb, 88.3%$^{207}$Pb, 9.06% $^{208}$Pb); $^{206}$Pb (94.0%$^{206}$Pb, 4.04%$^{207}$Pb, 1.96% $^{208}$Pb); $^{nat}$Pb (1.4%$^{204}$Pb, 24.1%$^{206}$Pb, 22.1%$^{207}$Pb, 52.4% $^{208}$Pb); $^{209}$Bi>99.9%. The $^{27}$Al(p,x)$^{22}$Na reaction was used for monitoring the proton flux. The proton fluencies were from $3.1*10^{13}$ to $1.4*10^{14}$ p/cm$^2$. The produced radionuclides were detected using the direct gamma spectrometry method with a Ge detector of 1.8 keV resolution at the 1332 keV $^{60}$Co gamma line. Each of the irradiated targets was measured during 3 to 6 months. The gamma spectra were processed using the interactive mode of the GENIE200 program using preliminary results from automatic-mode processing. The results of gamma-spectra processing serve as input to the SIGMA code that identifies the measured gamma lines using the PCNUDAT nuclear decay data base and determines the cross sections of the found radionuclides. Fig.1 shows the distributions of experimental uncertainties. Details of the experimental techniques are described in [2].

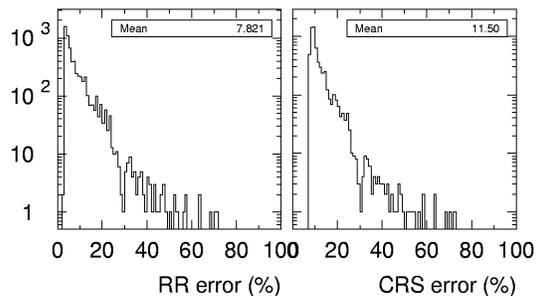

Fig. 1. Statistics of Reaction Rates (left panel) and cross section (right panel) uncertainties.

In total, 5972 nuclide production cross sections were measured in 55 experiments. The data themselves and their graphical interpretation are presented in the final technical report of the ISTC Project #2002 and will be uploaded into the EXFOR data base.

## 3. Theoretical Modeling

Seven codes were used to simulate the measured cross sections: LAHET (Bertini and ISABEL options) [5], the 2003 versions of CEM2k+GEM2 (CEM03) and of LAQGSM+GEM2 (LAQGSM03) [6], INCL4+ABLA [7, 8], CASCADE (old and the 2004 versions) [9], CASCADO and LAHETO [10]. The two latter codes are recent IPPE modifications of the CASCADE and LAHET codes, respectively. 884 figures with excitation functions (EF) by the seven codes and experimental data (ED) have been drawn. As an example, some of those figures are presented here in Figs. 2 and 3. For a quantitative comparison, we use the mean simulated-to-experiment squared deviation factor <F>, as described in [2].

To understand how different codes agree with the data in different nuclide production regions, we divided conventionally all products into four groups: shallow spallation products (A>170), deep spallation products (140<A<170), fission products (30<A<140), and fragmentation products (A<30). Besides, the energy ranges are conditionally broken into groups of low ($E_p$<0.1GeV), medium (0.1GeV<$E_p$<1.0GeV), and high ($E_p$>1.0GeV) energies. The mean deviation factors <F> are presented in Table 1 for each group separately together with the <F> values for all the comparison events. To ease the reading, the three lowest values of <F> are shown in red, while the three highest values in blue, for each of the comparison group.

- **A>170 (shallow spallation products):** most of the products from this region are predicted satisfactorily, with a mean deviation factor below 2. The near-target products (A above 200) are predicted variously at different proton energies: For instance, CEM03 predicts such products with <F>~1.5 at energies below 1 GeV, but underestimates them significantly (<F>~6) at energies above 1 GeV. On the contrary, LAHET and LAQGSM03 predict these products with <F>~1.5-2 at energies above 1 GeV, but fail to do so well at lower energies (<F>~4-5). The same is true for INCL4+ABLA: <F>~1.3-1.5 at $E_p$>0.1GeV and <F>~6 at $E_p$<0.1GeV. As all codes provide similar <F> values averaged over all energies, it is difficult to choose the best among them.

- **140<A<170 (deep spallation products):** With decreasing the mass of products, the predictive power of almost all codes also decreases. The degradation of the predictive power of different codes varies. For example, for BERTINI, <F> increases up to only 1.9; for LAQGSM03, <F> increases up to 2.3; and in the case of INCL4+ABLA, <F> increases up to 3.7. The INCL4+ABLA underestimates significantly the deep spallation products, thus overestimating their threshold energies. Judging from the <F> values, CASCADE-2004 is much ahead of other codes (<F> =1.47 against <F> =1.81 for BERTINI) in this region.

- **Fission products (FP)** are about a third of all measured and analyzed nuclides, and are described by the codes worse than the spallation products. INCL4+ABLA and CEM03, as well as LAHETO, show the best predictive power for fission products, whose <F> is ranging from 2.0 to 2.3. A peculiar agreement is demonstrated by the code INCL4+ABLA: <F> is too high (up to 6) in the 120<A<140 region where FP's overlap with deep spallation products, however, its agreement becomes the best (<F> from 1.5 to 2.0) in comparison with other codes for fission products with A<120. LAQGSM03 shows somewhat bigger deviation from ED (<F> up to 4), however, the agreement is better in the 80<A<110 region, with <F> around 2. CASCADE shows a much worse agreement on FP's (<F> up to ~20) compared with other codes.

- **Fragmentation products** are much underestimated by all the codes tested. The simulations underestimate the measured fragmentation yields by an order and more. As a whole, the CEM03 and LAQGSM03 results for these fragments are closest to ED.

Finally, we like to mention that as the gamma-spectrometry method used to obtain only part of the products, our comparison does not pretend to be comprehensive and to choose the best from the tested codes. Rather, it points to some distinct problems each code still has, helping the authors of the codes to further improve them.

Table 1. Mean squared deviation factors <F> separately for different energy groups and ranges of products (A>30) and for all comparisons as well.

| Code | Product mass (A) | | | Proton energy ($E_p$, GeV) | | | Total |
|---|---|---|---|---|---|---|---|
| | A>170 | 140<A<170 | 30<A<140 | $E_p$<0.1 | 0.1<$E_p$<1.0 | $E_p$>1.0 | |
| ISABEL | 1.81 | 1.81 | 2.87 | 4.88 | 2.13 | - | 2.16 |
| BERTINI | 1.75 | 1.93 | 2.75 | 4.26 | 2.06 | 1.97 | 2.10 |
| INCL4+ABLA | 1.90 | 3.74 | 2.22 | 4.63 | 2.18 | 2.13 | 2.25 |
| CASCADE | 1.77 | 2.01 | 6.93 | 4.93 | 3.93 | 2.44 | 3.25 |
| CASCADE-2004 | 1.93 | 1.47 | 5.54 | 6.54 | 3.23 | 2.42 | 2.94 |
| LAQGSM03 | 1.98 | 2.32 | 2.71 | 3.03 | 2.35 | 2.09 | 2.26 |
| CEM03 | 1.98 | 2.07 | 2.25 | 2.08 | 1.77 | 2.39 | 2.07 |
| CASCADO | 1.99 | 2.22 | 2.83 | 2.69 | 2.33 | 2.22 | 2.29 |
| LAHETO | 1.99 | 1.96 | 1.98 | 4.85 | 1.76 | - | 1.98 |

## 4. Conclusions

5972 product cross sections are measured in 55 experiments at ITEP and analyzed with seven codes. The predictive powers of the codes tested here vary but were found to be satisfactory for most of the nuclides in the spallation region, though none of the benchmarked codes agrees well with all the data in the whole mass region of product nuclides and all codes should be improved further. On the whole, the predictive power of all codes for the data in the fission product region is worse than in the spallation region; the agreement is even worse in the fragmentation region and on the border between spallation and fission regions; development of better evaporation, fission, and fragmentation models is of high priority.

## Acknowledgments

The work has been performed under the ISTC Project #2002 supported by the European Community. It was partially supported by the Advanced Simulation Computing (ASC) Program at the Los Alamos Na-tional Laboratory operated by the University of California for the U. S. Department of Energy and the NASA Grant NRA-01-ATP-066.

✚ – [4], ○ - [3]) ( LAHET –black (ISABEL - solid, BERTINI – dashed), CEM03 - magenta, INCL4+ABLA – red, CASCADE – green, LAQGSM03 - green-blue, LEHETO- blue,  CASCADO – dashed blue).

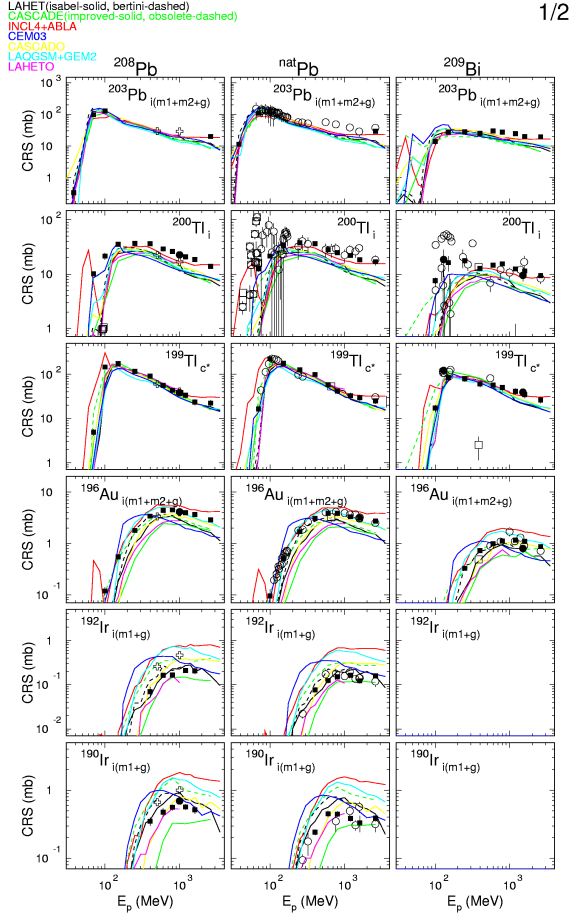

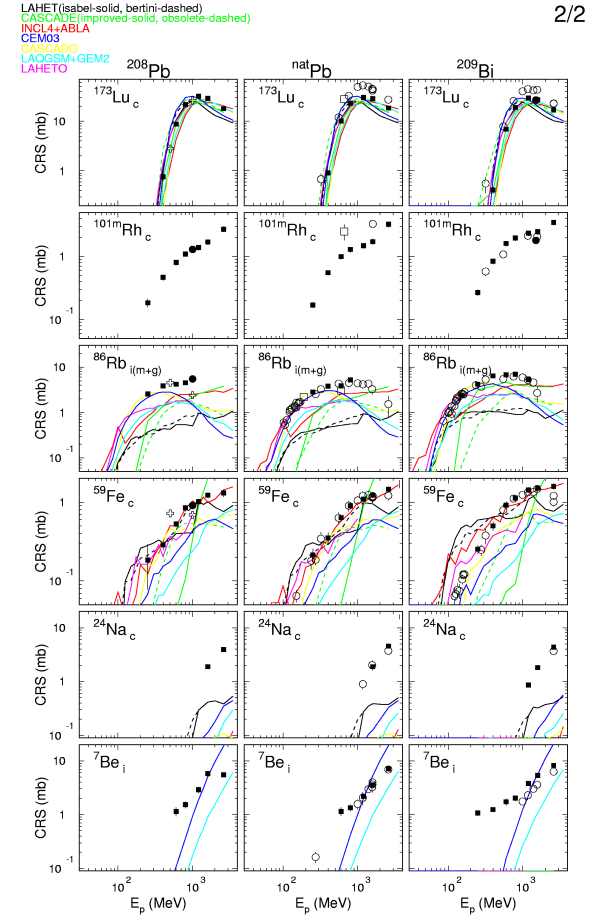

**Fig. 2.** Experimental and simulated excitation functions of $^{203}$Pb, $^{200}$Tl, $^{199}$Tl, $^{196}$Au, $^{192}$Ir, and $^{190}$Ir produced in $^{208}$Pb (left), $^{nat}$Pb (center), and $^{209}$Bi(right). (■ – this work, ● – [2],

**Fig. 3.** The same as Fig. 2 but for $^{173}$Lu, $^{101m}$Rh, $^{86}$Rb, $^{59}$Fe, $^{24}$Na, and $^{7}$Be.